\documentstyle[11pt]{article}
\newcommand{\be}{\begin{equation}}
\newcommand{\ee}{\end{equation}}

\newcommand{\bea}{\begin{eqnarray}}
\newcommand{\eea}{\end{eqnarray}}
\newcommand{\mbold}[1]{\mbox{\boldmath$#1$}}

\begin{document}

\title{One-particle exchange in the double folded potential in
a semiclassical approximation}
\author{V.B.Soubbotin \\
{\small Nuclear Physics Department, Physical Research Institute},\\
{\small St.Petersburg University, St.Petersburg, Russia} \\
{X.Vi\~nas}\\
{\small Departament d'Estructura i Constituents de la Mat\`eria, Facultat de
F\'{\i}sica,} \\
{\small Universitat de Barcelona, Diagonal 647 E-08028 Barcelona, Spain }}
\maketitle

\begin{abstract}
The one-particle exchange in the double folded model is analyzed. To this
aim the Extended Thomas-Fermi approach to the one-body density matrix is
used. The nucleon- nucleon force with Yukawa, Gauss and
Coulomb-type form factors are considered.
The energy dependence of the exchange part of
the double folded potential is investigated and a comparison of the
present approach with former ones is carried out.
\end{abstract}

\pagebreak

\section{Introduction}

Up to the present moment the Double Folded Model (DFM) 
\cite{SL,Sinh,Kh}, which starts
from the effective nucleon-nucleon force and the single particle densities
of the colliding nuclei, has become one of the most popular methods to
calculate the real part of microscopic the Heavy-Ion (HI) optical
potential. It
is also known that folding methods are widely used in other problems of HI
Physics \cite{Kr}. While in the original DFM the antisymmetrization effects
were taken into account by means of the effective zero-range pseudopotential 
, in the latest versions of the DFM the antisymmetrization has been
explicitly considered \cite{SL,Sinh,Kh}. It is assumed that the structure of
of each isolated nucleus is described at Hartree-Fock (HF) level
and that the exchange
effects come from the non-diagonal part of the single-particle density
matrix (DM). The intrinsic states of isolated nuclei are modified during the
interaction. However in the DFM the so called ''frozen density
approximation'', which implies
that the intrinsic states of the nuclei do not change during the
interaction, is assumed.

Thus the key point in the DFM is to express the non-diagonal part of the
DM through its diagonal part or, in other words, in a Local Density
Approximation (LDA) for DM. The simplest LDA is the Slater (SL) approach to
the DM, which is equivalent to the Thomas-Fermi approach, valid in
nuclear matter, and does not take into account the nuclear surface effects.
A more elaborated LDA is given by the Density Matrix Expansion (DME) of
Negele and Vautherin (NV) \cite{NV}. Another approach widely used in
DFM \cite{Kh,Kn} is the Campi-Bouassy (CB) \cite{CB} approximation, in which
by a proper choice of the effective momenta in the DME of NV, reduces to the
SL form.

The physical quantities that enter into the NV and CB approaches are just the
Fermi momentum $k_F$ and the kinetic energy density $\tau $. In the HF
method the single-particle density $\rho $ as well as $k_F$ and $\tau $ are
determined selfconsistently. However, the attractive feature of DFM is to
consider $\rho $ as an input data. While the densities $\rho $ can be
obtained from the electron scattering data, $k_F$ and $\tau $ should be
defined theoretically. At present, the values of $k_F$ and $\tau $ that
correspond to a given single-particle density $\rho $ are unknown. To
determine them via $\rho $, approximate schemes are in order. In particular
the Extended Thomas-Fermi (ETF) approach is widely used to this end: 
\begin{equation}
k_F=\bigg( \frac{3\pi ^2\rho }2\bigg)^{1/3}  \label{eq1}
\end{equation}
and 
\begin{equation}
\tau =\frac 35(\frac{3\pi ^2}2)^{2/3}\rho ^{5/3}+\frac 1{36}\frac{(\nabla
\rho )^2}\rho +\frac 13\Delta \rho  \label{eq2}
\end{equation}
if a degeneracy 4 is assumed.

Very recently the ETF approximation has been used for deriving the
semiclassical one-body density matrix up to $\hbar ^2$ order in the case of
a non-local single-particle potential \cite{SV}. This ETF DM is obtained
starting from the Wigner-Kirkwood distribution function for a non-local
one-body Hamiltonian \cite{Cen}. As is shown in \cite{SV}, the NV and CB
approaches to the DM with $\tau $ calculated in the ETF approach, are just
truncations of the full ETF DM. Consequently, it seems appealing
to use the complete
semiclassical DM to obtain the real part of the ion-ion potential and
compare these results with former potentials obtained using these
semiclassical NV and CB approximations.

In the present paper we apply the ETF DM \cite{SV} for calculating the DFM potential.
We obtain analytical expressions for the potential using the Gogny \cite{Gog},
M3Y \cite{SL} and Coulomb forces. To check the validity of our ETF
approximation, we analyze the $^{16}O$- $^{16}O$ reaction. The ground state
of each isolated $^{16}O$ nucleus is described using harmonic oscillator
(HO) wavefunctions. In this case the quantal DFM potential can be calculated
explicitly \cite{Ism} and we compare this quantal potential with the semiclassical DFM
potential obtained using the ETF, SL and CB approximations to the DM.

The paper is organized as follows. In the first section we present the
analytical derivation of the semiclassical DFM potential using  the SL,
CB and ETF approximations to the DM and Gauss, Yukawa and Coulomb
form factors for the nucleon-nucleon force. In the next section we compare
these semiclassical DFM nucleus-nucleus potentials obtained
using several prescriptions for the density of eachisolated nucleus with
the corresponding quantal potential. This comparison is done for the $^{16}O$-$^{16}O$
reactions. The summary and conclusions are given in the last section.

\section{Formalism}

In order to obtain the HI potential in the DFM, one can start from its microscopic
definition 
\begin{equation}
V(D)=\sum_{i\in 1,j\in 2}[<ij|v|ij>-<ij|v|ji>],  \label{eq3}
\end{equation}
where $i$ and $j$ represent single particle states of the first- and
second-nuclei respectively and $v$ is the effective nucleon-nucleon force.
Formally (\ref{eq3}) coincides with the corresponding term in the HF method.
However in the DFM approach, the single-particle wave functions $|i>$ and $|j>$
are determined selfconsistently only inside each nucleus independently. To
take into account the relative motion in the plane wave approximation one should
multiply $|i>$ by a factor $exp(i\mbox{\boldmath$k$}_1\mbox{\boldmath$r_i$})$
in the first nucleus and $|j>$ by $exp(i\mbox{\boldmath$k$}_2%
\mbox{\boldmath$r$}_j)$ in the second. Here $\mbox{\boldmath$k$}_1=%
\mbox{\boldmath$K$}_1/A_1$, $\mbox{\boldmath$k$}_2=\mbox{\boldmath$K$}_2/A_2$
and $\mbox{\boldmath$K$}_\alpha (\alpha =1,2)$ is the centre-of-mass
moment
of the first and second nucleus respectively. The relative momentum between
a pair of nucleons of different nuclei is given by : $\mbox{\boldmath$K$}=%
\mbox{\boldmath$k$}_1-\mbox{\boldmath$k$}_2$. Its modulus can be defined
globally as $K^2=2m \mu E_{cm}/ \hbar ^2$ or
locally as \cite{Sinh}: $K^2=2m \mu (E-V-V_{Coul})/\hbar ^2$,
where $\mu =A_1A_2/(A_1+A_2)$. In the local definition
the set of equations: $V=V(K);K=K(V)$ should be solved selfconsistently.

The direct part of the HI potential is given by the first matrix element of (%
\ref{eq3}) and reads: 
\begin{eqnarray}
V_{dir}(D)=X_d\int d\vec{r}_1d\vec{r}_2 \rho_1(\vec{r}_1) \rho_2(\vec{r}_2-%
\vec{D})v(s), \label{eq4}
\end{eqnarray}
where $\mbox{\boldmath$s$}=\mbox{\boldmath$r$}_1-\mbox{\boldmath$r$}_2$. The
direct part of the HI potential eq.(\ref{eq4}) is just the convolution of
the single-particle densities $\rho_i$ with the form factor $v(s)$ of the
central effective nucleon-nucleon force $v$. The constant $X_d$ is just the
standard combination of the exchange parameters of the central
nucleon-nucleon force: $X_d=w+b/2-h/2-m/4$. This contribution is easily
calculated numerically. It is important to note that the direct part of
the HI potential \ref{eq4} does not depend on energy if the nucleon-nucleon
interaction $v$ is energy independent.

The finite-range exchange part of the HI potential is given by: 
\begin{equation}
V_{ex}(D,K)=X_e\int d\vec r_1d\vec r_2\rho _1(\vec r_1,\vec r_2)\rho _2(\vec
r_2-\vec D,\vec r_1-\vec D)v(s)e^{i\mbox{\boldmath$K$}\mbox{\boldmath$s$}}
. \label{eq5}
\end{equation}
Here, as usual, $X_e=m+h/2-b/2-w/4$. To obtain this term a LDA for the
density matrix is used. The simplest LDA is the SL approach, which in
terms of centre-of-mass $\mbox{\boldmath$R$}=(\mbox{\boldmath$r_1$}+%
\mbox{\boldmath$r_2$})/2$ and relative $\mbox{\boldmath$s$}$ coordinates
reads as: 
\begin{equation}
\rho (\vec r_1,\vec r_2)=\rho (\mbox{\boldmath$R$})\frac{3j_1(k_Fs)}{k_Fs}
, \label{eq6}
\end{equation}
where the Fermi momentum $k_F$ is related to the density by the usual
Thomas-Fermi relation (\ref{eq1}) (Here we use degeneracy factor 4 so the
single particle density $\rho $ is normalized to the mass number). This
approximation is exact at HF level in infinite nuclear matter. To take into
account finite size effects of the nuclei some modifications have to be
made. Among different approaches to the DM, in the present paper we consider
the Campi-Bouyssy (CB) \cite{CB} one that is often used in DFM calculations.
In the CB approximation the DM has the same form as the SL term (\ref{eq3}),
but instead of the Fermi momentum $k_F$,
uses an effective momentum $\tilde k$ defined as: 
\begin{equation}
\tilde k^2=\frac 5{3\rho (\mbox{\boldmath$R$})}(\tau (\mbox{\boldmath$R$}%
)-\frac 14\Delta \rho (\mbox{\boldmath$R$})).  \label{eq7}
\end{equation}
In this equation $\tau $ is just the exact quantal kinetic energy density.
Consequently, to calculate it one needs to know the exact DM. To overcome
this difficulty some approximate schemes for $\tau $ are used, often they
are based on its ETF value eq.(\ref{eq2}) \cite{Ism}.

Let us now derive the exchange term (\ref{eq5}) using the ETF approach to
the DM \cite{SV} for each isolated nucleus.
This DM averaged over $\mbox{\boldmath$s$}$ can be written as: 
\begin{equation}
\rho (\vec R,s)=\rho _{TF}(\mbox{\boldmath$R$},s)+\delta \rho (%
\mbox{\boldmath$R$},s), \label{eq8}
\end{equation}
where $\rho _{TF}$ is the Thomas-Fermi term, which is equivalent to the SL
approximation (\ref{eq6}). The averaged second-order semiclassical
correction $\delta \rho $ that we will use in this work reads as: 
\begin{equation}
\delta \rho =\frac{s^2}{72}\{\Delta \rho [j_0(k_Fs)-\frac{6j_1(k_Fs)}{k_Fs}]-%
\frac{(\nabla \rho )^2}\rho [\frac 43j_0(k_Fs)-3\frac{j_1(k_Fs)}{k_Fs}]\}.
\label{eq9}
\end{equation}
In this approximation all the quantities are defined in a unique way. Eq.(%
\ref{eq9}) gives the ETF-$\hbar ^2$ correction to the DM in the case of a
local HF potential. To take into account the non-locality of the single
particle mean field, one should consider effective masscorrections in
$\delta \rho$. This case is analyzed in \cite{SV} in detail. It is also shown
that the contribution of these non-local effects to the $\hbar ^2$ part of
the DM are rather small as compared with the full $\hbar ^2$ correction if
realistic finite-range forces are used. Therefore we will not consider the
effective mass corrections to $\delta \rho$ in this paper.
We calculate the DFM potential using nucleon-nucleonforces
with a Gauss, Yukawa and Coulomb-type form factors which cover most of
the effective interactions used in DFM calculations. As typical examples we will
consider the Gogny \cite{Gog} and M3Y forces \cite{SL}.

\subsection{Gaussian-type force}

Let us consider a central nucleon-nucleon force with a Gaussian form factor: 
\begin{equation}
v(s)=v_0exp(-\frac{s^2}{\mu ^2}).  \label{eq10}
\end{equation}
The exchange contribution to the DFM potential is given by:
\begin{equation}
V_{ex}(D)=X_ev_0\int d^3r\{\tilde V_{Sl}(k[r])+\delta \tilde V(k[r])\},
\label{eq11}
\end{equation}
where the SL term $\tilde V_{Sl}$ is given by 
\begin{eqnarray}
\tilde V_{Sl} &=&-\frac{9\pi ^2\rho _1\rho _2}{2Kk_1^3k_2^3}\sum_\alpha
\{[-\sigma _1\sigma _2(\frac{\alpha ^4}{24}+\frac{\alpha ^2}{2\mu ^2}+\frac
1{2\mu ^4}) \nonumber \\
&&+(\sigma _2k_1+\sigma _1k_2)(\frac{\alpha ^3}6+\frac \alpha {\mu
^2})-k_1k_2(\frac{\alpha ^2}2+\frac 1{\mu ^2})]erf(\frac \mu 2\alpha ) 
\nonumber \\
&&+[-\sigma _1\sigma _2(\frac{\alpha ^3}{12\mu ^2}+\frac{5\alpha }{6\mu ^4}%
)+(\sigma _2k_1+\sigma _1k_2)(\frac{\alpha ^2}{3\mu ^2}+\frac 4{3\mu ^4}) 
\nonumber \\
&&-k_1k_2\frac \alpha {\mu ^2}]\frac \mu {\sqrt{\pi }}exp[-\frac{\mu ^2}%
4\alpha ^2]\}.  \label{eq12}
\end{eqnarray}
Here $\sigma _{1,2}$ is the sign of $k_{1,2}$ (which are the Fermi momentum $%
k_F$ of each nucleus related to the corresponding density through eq.(\ref
{eq1})), while $\alpha =K\pm k_1\pm k_2$. The summation in (\ref{eq12}) is
taken over all possible $\alpha $. The explicit form of (\ref{eq12}) is
given in \cite{GSFV}. To calculate the DFM potential in the CB approach we
use the effective momentum (\ref{eq7}) with the semiclassical ETF kinetic
energy density (\ref{eq2}).

The second-order correction to the DFM is given by:
\begin{eqnarray}
\delta \tilde V(R) &=&-\frac 1{36\pi ^2K}\{(-k_1f_1-k_2f_2)\sum_\alpha
\sigma _1\sigma _2erf(\frac \mu 2\alpha ) \nonumber \\
&&+\frac{\mu k_1k_2}{\sqrt{\pi }}[(f_1+f_2)(exp[-\frac{\mu ^2(K+x_1)^2}%
4]-exp[-\frac{\mu ^2(K-x_1)^2}4])  \nonumber \\
&&+(f_1-f_2)(exp[-\frac{\mu ^2(K+x_2)^2}4]-exp[-\frac{\mu ^2(K-x_2)^2}4])] 
\nonumber \\
&&-3(\frac{g_1}{k_1}+\frac{g_2}{k_2})[[\frac 12(K^2-k_1^2-k_2^2)+\frac 1{\mu
^2}]\sum_\alpha (\sigma _1)(\sigma _2)erf[\frac{\mu \alpha }2]  \nonumber \\
&&+\frac 1{\sqrt{\pi }\mu }\sum_\alpha \sigma _1\sigma _2\alpha ^{-}exp[-%
\frac{\mu ^2}4\alpha ^2]]\},  \label{eq13}
\end{eqnarray}
where $\alpha ^{-}=K\mp k_1\mp k_2$, $x_1=k_1+k_2$, $x_2=k_1-k_2$ and 
\begin{eqnarray}
f_i &=&\Delta k_i-2\frac{(\nabla k_i)^2}{k_i}=\frac{\pi ^2}{2k_i^2}[\Delta
\rho _i-\frac 43\frac{(\nabla \rho _i)^2}{\rho _i}] \\
g_i &=&2\Delta k_i+\frac{(\nabla k_i)^2}{k_i}=\frac{\pi ^2}{2k_i^2}[2\Delta
\rho _i-\frac{(\nabla \rho _i)^2}{\rho _i}].  \nonumber  \label{eq14}
\end{eqnarray}
Let us now analyze these formulas.
We will talk about the high energy limit when 
$K/k_i>>1,i=1,2$ which is valid at high enough energy. However, it could also
be valid at small energies in the outer region of nuclei when they overlap
weakly. One can see that in this regime the SL term dominates and the $h^2$
correction is negligible. 
\begin{equation}
\tilde V_{Sl}=-\frac{3\pi ^{3/2}\rho _1\rho _2}{8k_1^3k_2^3}\mu
^3K^4k_1k_2e^{-\frac{\mu ^2K^2}4}+O([\frac{k_i}K]^2]).  \label{eq15}
\end{equation}
On the contrary, at small energies the $h^2$ correction plays an important
role. The SL term in the low energy limit, e.g. when $\frac K{k_i}<<1$,
looks like 
\begin{eqnarray}
\tilde V_{Sl} &=&-\frac{9\pi ^2\rho _1\rho _2}{2k_1^3k_2^3}\{\frac
23(k_1^3+k_2^3)erf[\frac \mu 2(k_1+k_2)]-\frac 23(k_1^3-k_2^3)erf[\frac \mu
2(k_1-k_2)] \nonumber \\
&&+\frac 1{\mu \pi }[\frac 8{3\mu ^2}(exp[-\frac{\mu ^2}4(k_1-k_2)^2]-exp[-%
\frac{\mu ^2}4(k_1+k_2)^2])  \nonumber \\
&&+\frac 43(k_1+k_2)^2exp[(-\frac{\mu ^2}4(k_1+k_2)^2]-\frac
43(k_1-k_2)^2exp[(-\frac{\mu ^2}4(k_1-k_2)^2]  \nonumber \\
&&-4k_1k_2(exp[-\frac{\mu ^2}4(k_1-k_2)^2]+exp[-\frac{\mu ^2}%
4(k_1+k_2)^2])]\}+O([\frac K{k_i}]^2).  \label{eq16}
\end{eqnarray}
This formula was obtained in \cite{JP} for the static SL DFM potential with
the Gogny force. The low energy limit of the $\hbar ^2$ correction is: 
\begin{eqnarray}
\delta \tilde V(R) &=&-\frac \mu {36\pi ^{5/2}}\{[3(\frac{g_1}{k_1}+\frac{g_2%
}{k_2})[(k_1^2+k_2^2)-\frac 4{\mu ^2}]-2(k_1f_1+k_2f_2)] \nonumber \\
&&\times [exp[-\frac{\mu ^2}4(k_1+k_2)^2]-exp[-\frac{\mu ^2}4(k_1-k_2)^2]] 
\nonumber \\
&&-3(\frac{g_1}{k_1}+\frac{g_2}{k_2})[(k_1+k_2)^2\exp [-\frac{\mu ^2}%
4(k_1+k_2)^2]  \nonumber \\
&&-(k_1-k_2)^2\exp [-\frac{\mu ^2}4(k_1-k_2)^2]] \nonumber \\
&&-\mu ^2k_1k_2[(k_1+k_2)(f_1+f_2)exp[-\frac{\mu ^2}4(k_1+k_2)^2]  \nonumber
\\
&&+(k_1-k_2)(f_1-f_2)exp[-\frac{\mu ^2}4(k_1-k_2)^2]]\}+O((\frac K{k_i})^2)
. \label{eq17}
\end{eqnarray}
In a realistic calculation both limiting cases: $K/k_i<<1$ and $k_i/K<<1$
could take place at the same relative energy, but at different points in
space.

\subsection{Yukawa-type force}

For a Yukawa-type force 
\begin{equation}
v(s)=v_0\frac{e^{-\beta s}}{\beta s}  \label{eq18}
\end{equation}
one will get 
\begin{eqnarray}
\tilde V_{Sl} &=&-\frac{9\pi \rho _1\rho _2}{K\beta k_1^3k_2^3}\{\frac
1{30}K^3k_1k_2+\frac{11}{30}(k_1^2+k_2^2)Kk_1k_2-\frac 1{10}\beta ^2Kk_1k_2
\nonumber \\
&&+\sum [-\sigma _1\sigma _2(\frac{\alpha ^2\beta ^3}{12}-\frac{\alpha
^4\beta }{24}-\frac{\beta ^5}{120})  \nonumber \\
&&+(\sigma _2k_1+\sigma _1k_2)(\frac{\alpha \beta ^3}6-\frac{\alpha ^3\beta }%
6)+k_1k_2(\frac{\alpha ^2\beta }2-\frac{\beta ^3}6)]arctg(\frac \alpha \beta
)  \nonumber \\
&&+[-\sigma _1\sigma _2(\frac{\alpha ^5}{120}+\frac{\alpha \beta ^4}{24}-%
\frac{\alpha ^3\beta ^2}{12})  \nonumber \\
&&+(\sigma _2k_1+\sigma _1k_2)(\frac{\alpha ^4}{24}-\frac{\alpha ^2\beta ^2}%
4+\frac{\beta ^4}{24})+k_1k_2(\frac{\alpha \beta ^2}2-\frac{\alpha ^3}%
6)]J_1^c(\alpha )\},  \label{eq19}
\end{eqnarray}
where only the combination 
\[
J_1^c(\alpha _1)-J_1^c(\alpha _2)=\int \frac{ds}s[cos(\alpha _1s)-cos(\alpha
_2s)]exp(-\beta s)=\frac 12ln\frac{\alpha _2^2+\beta ^2}{\alpha _1^2+\beta ^2%
} 
\]
appears. Again if one uses the CB effective momenta one will get the CB
approximation to the DFM. The explicit expression of (\ref{eq19}) can be
found in \cite{GSFV}. The second-order correction to the DFM potential is: 
\begin{eqnarray}
\delta \tilde V &=&-\frac 1{18\pi ^3\beta K}\{2Kk_1k_2(\frac{g_1}{k_1}+\frac{%
g_2}{k_2}) \\
&&+[\beta (k_1f_1+k_2f_2)-(\frac{\beta ^3}2+\frac{3\beta }2(k_1^2+k_2^2)-%
\frac{3\beta }2K^2)(\frac{g_1}{k_1}+\frac{g_2}{k_2})]\sum_\alpha \sigma
_1\sigma _2arctg(\frac \alpha \beta )  \nonumber \\
&&+[\frac K2(k_1f_1+k_2f_2)+(\frac{K^3}4-\frac{3K}4(k_1^2+k_2^2+\beta ^2))(%
\frac{g_1}{k_1}+\frac{g_2}{k_2})  \nonumber \\
&&\times [ln\frac{(K+x_1)^2+\beta ^2}{(K+x_2)^2+\beta ^2}+ln\frac{%
(K-x_1)^2+\beta ^2}{(K-x_2)^2+\beta ^2}]  \nonumber \\
&&+[\frac 12(k_1^2f_1+k_2^2f_2)-\frac 12(k_1^3+k_2^3)(\frac{g_1}{k_1}+\frac{%
g_2}{k_2})]ln\frac{(K+x_1)^2+\beta ^2}{(K-x_1)^2+\beta ^2}  \nonumber \\
&&-[\frac 12(k_1^2f_1-k_2^2f_2)-\frac 12(k_1^3-k_2^3)(\frac{g_1}{k_1}+\frac{%
g_2}{k_2})]ln\frac{(K+x_2)^2+\beta ^2}{(K-x_2)^2+\beta ^2}\}. \label{eq20}
\end{eqnarray}
Considering the high energy limit, where the $\hbar ^2$ correction to the
potential is negligible, one will get 
\begin{equation}
\tilde V_{Sl}=-\frac{4\pi }{\beta K^2}\rho _1(\vec r)\rho _2(\vec r-\vec
R)+O([\frac{k_i}K]^4).  \label{eq21}
\end{equation}
These results for a Yukawa force correspond to the convolution of the
densities of the isolated nuclei with a zero-range pseudopotential and an
extra $1/E$ energy-dependence. Therefore at high enough energy, when $K/k_i>1$%
, a simple zero-range pseudopotential can be used for calculating the
exchange part of the DFM potential in this case. The parameters of this pseudopotential
are determined by the range $\beta $ of the effective nucleon-nucleon force
(here M3Y). The inverse energy dependence in (\ref{eq21}) should be taken into
account.

Now we consider the low energy limit where $h^2$ correction is more
important. The SL term in this case looks like: 
\begin{eqnarray}
\tilde V_{Sl} &=&-\frac{9\pi \rho _1\rho _2}{\beta k_1^3k_2^3}\{k_1k_2
\lbrack \frac{11}{30}(k_1^2+k_2^2)-\frac 1{10}\beta ^2 \\
&+&\frac 1Q[\frac{16\beta ^2}{15}k_1^2k_2^2+\frac{\beta ^2}3(k_1^2-k_2^2)^2-%
\frac{\beta ^6}{15}+\frac 2{15}(k_1^2-k_2^2)^2(k_1^2+k_2^2-\beta ^2)]\rbrack
\nonumber \\
&-&\frac{2\beta }3[(k_1^3+k_2^3)arctg(\frac{k_1+k_2}\beta
)-(k_1^3-k_2^3)arctg(\frac{k_1-k_2}\beta) ]  \nonumber \\
&+&(\frac{\beta ^4}{24}+\frac{\beta ^2}4(k_1^2+k_2^2)-\frac
18(k_1^2-k_2^2)^2)\ln \frac{(k_1+k_2)^2+\beta ^2}{(k_1-k_2)^2+\beta ^2}%
\}+O((\frac K{k_i})^2).  \nonumber  \label{eq22}
\end{eqnarray}
The corresponding $\hbar ^2$ correction reads: 
\begin{eqnarray}
\delta \tilde V &=&-\frac 1{9\pi ^3\beta }\{3k_1k_2(\frac{g_1}{k_1}+\frac{g_2%
}{k_2}) \\
&-&\frac{2k_1k_2}Q[\beta ^2(k_1f_1+k_2f_2)+(k_1^2-k_2^2)(k_1f_1-k_2f_2)] 
\nonumber \\
&+&\frac 12[(k_1f_1+k_2f_2)-\frac 32(k_1^2+k_2^2+\beta ^2)(\frac{g_1}{k_1}+%
\frac{g_2}{k_2})]ln\frac{(k_1+k_2)^2+\beta ^2}{(k_1-k_2)^2+\beta ^2}%
\}+O([\frac K{k_i}]^2),  \nonumber  \label{eq23}
\end{eqnarray}
where 
\begin{equation}
Q=\beta ^4+2\beta ^2(k_1^2+k_2^2)+(k_1^2-k_2^2)^2  \label{eq23a}
\end{equation}

\subsection{Coulomb force}

Let us now consider the Coulomb force: 
\begin{equation}
v(s)=\frac{e^2}r  \label{eq24}
\end{equation}
In this particular case one will get the explicit expression for the SL
term: 
\begin{eqnarray}
\tilde V_{Sl} &=&-\frac{9\pi e^2\rho _1\rho _2}{Kk_1^3k_2^3}\{\frac
1{30}K^3k_1k_2+\frac{11}{30}(k_1^2+k_2^2)Kk_1k_2 \nonumber \\
&&+[\frac{K^5}{240}-\frac{K^3}{24}(k_1^2+k_2^2)-\frac
K{16}(k_1^2-k_2^2)^2][ln\frac{(K+k_1+k_2)^2}{(K+k_1-k_2)^2}+ln\frac{%
(K-k_1-k_2)^2}{(K-k_1+k_2)^2}]  \nonumber \\
&&-[\frac{K^2}{12}(k_1^3+k_2^3)+\frac 1{60}(k_1+k_2)^3[(k_1-k_2)^2-k_1k_2]]ln%
\frac{(K+k_1+k_2)^2}{(K-k_1-k_2)^2}  \nonumber \\
&&+[\frac{K^2}{12}(k_1^3-k_2^3)+\frac 1{60}(k_1-k_2)^3[(k_1+k_2)^2+k_1k_2]]ln%
\frac{(K+k_1-k_2)^2}{(K-k_1+k_2)^2}\},  \label{eq25}
\end{eqnarray}
where now $\rho_i$ and $k_i$ (i=1,2) correspond to the density
and Fermi momentum of protons in each nucleus:
$\rho_i= \rho_{i,p}$, $k_i=k_{i,p}$. On first sight of (\ref{eq23a})
one can think that
at $K=\pm k_1\pm k_2$ this formula contains divergent terms. However, one
can see that the corresponding terms that appears twice cancel each other.

The second-order correction to the DFM potential is: 
\begin{eqnarray}
\delta \tilde V &=&-\frac{e^2}{18\pi ^3K}\{2Kk_1k_2(\frac{g_1}{k_1}+\frac{g_2%
}{k_2}) \nonumber \\
&&+[\frac K2(k_1f_1+k_2f_2)+(\frac{K^3}4-\frac{3K}4(k_1^2+k_2^2))(\frac{g_1}{%
k_1}+\frac{g_2}{k_2})  \nonumber \\
&&\times [ln\frac{(K+x_1)^2}{(K+x_2)^2}+ln\frac{(K-x_1)^2}{(K-x_2)^2}] 
\nonumber \\
&&+[\frac 12(k_1^2f_1+k_2^2f_2)-\frac 12(k_1^3+k_2^3)(\frac{g_1}{k_1}+\frac{%
g_2}{k_2})]ln\frac{(K+x_1)^2}{(K-x_1)^2}  \nonumber \\
&&-[\frac 12(k_1^2f_1-k_2^2f_2)-\frac 12(k_1^3-k_2^3)(\frac{g_1}{k_1}+\frac{%
g_2}{k_2})]ln\frac{(K+x_2)^2}{(K-x_2)^2}\}.  \label{eq26}
\end{eqnarray}
Considering the case $K=x_{1,2}$ one can see that (\ref{eq26}) still
remains well defined. In the high energy limit we get again the simple
formula:
\begin{equation}
\tilde V_{Sl}=-\frac{4e^2\pi }{K^2}\rho _1(\vec r)\rho _2(\vec r-\vec R)+O([%
\frac{k_i}K]^4)  \label{eq27}
\end{equation}
In the low-energy limit one obtains
\begin{equation}
\tilde V_{Sl}=-\frac{9\pi e^2\rho _1\rho _2}{k_1^3k_2^3}\{\frac
12k_1k_2(k_1^2+k_2^2)-\frac 18(k_1^2-k_2^2)^2ln\frac{(k_1+k_2)^2}{(k_1-k_2)^2%
}\}+O((\frac K{k_i})^2)  \label{eq28}
\end{equation}
and its $\hbar ^2$ correction reads as: 
\begin{eqnarray}
\delta \tilde V &=&-\frac{e^2}{9\pi ^3}\{3k_1k_2(\frac{g_1}{k_1}+\frac{g_2}{%
k_2})-\frac{2k_1k_2}{(k_1^2-k_2^2)}(k_1f_1-k_2f_2)  \nonumber \\
&&+\frac 12[(k_1f_1+k_2f_2)-\frac 32(k_1^2+k_2^2)(\frac{g_1}{k_1}+\frac{g_2}{%
k_2})]ln\frac{(k_1+k_2)^2}{(k_1-k_2)^2}\}+O([\frac K{k_i}]^2).
\label{eq29}
\end{eqnarray}

In the case $k_1=k_2$ one should take care about the correct consideration
of the limit $k_1\to k_2$. In this case one should start from eq.(26) for
the Yukawa force, put $k_1=k_2$ and make the limit $\beta lim_{\beta \rightarrow
0}$. After path integration one will get at $k_1=k_2$: 
\begin{equation}
\delta \tilde V=-\frac{7e^2}{18\pi }[(\nabla k_1)^2+(\nabla k_2)^2].
\label{eq30}
\end{equation}

It is very important to point out that the SL, CB or ETF approximations to
the DM described in this section are used for calculating the exchange part
of the DFM potential irrespective of the way that the density of each
isolated nucleus is obtained. In the following we will refer to these
potentials as SL, CB or ETF approaches to the DFM potential. Notice that for
determining completely the DFM potential, the density of each isolated
nucleus that enters through $k_1$ and $k_2$ in the previous formulas as well as the effective
nucleon-nucleon force used have to be specified.

\section{Results}

In this Section we present the results for the DFM potential obtained using
the different prescriptions considered in Section 2. The nuclear
part of these potentials calculated with the realistic effective
nucleon-nucleon force analyzed in this paper is attractive. Hence the
relative momentum $K$ in the local definition is always real when the energy
exceeds the Coulomb barrier. On the other hand, the ''frozen density''
approximation, used in DFM, is still valid for fast enough collisions or, in
other words, at rather high energies. The Coulomb barrier for the $^{16}O-^{16}O$
system is about 10MeV. Therefore, we will consider only collisions above
the Coulomb barrier where $K$ is always real.

To check the validity of the ETF DM to the calculation of an ion-ion potential, we
consider the exactly soluble HO model for the isolated nuclei. This model
consists of describing the ground state of each colliding nucleus using HO
wavefunctions. In this case the exchange part of the potential is easily
calculated due to factorization of variables in the DM \cite{Ism}.

First of all, let us briefly describe the effective nucleon-nucleon
force used in this paper. Here we will consider the Gogny force that
is used in nuclear structure calculations and a MY3-type force that is
typical in DFM calculations.

The effective Gogny force \cite{Gog} consists
of a finite-range Brink-Boeker term together with a density-dependent
zero-range term 
\begin{equation}
v(r)=\sum_{i=1}^2[w_i+b_iP^\sigma -h_iP^\tau -m_iP^\sigma P^\tau ]exp(-\frac{%
r^2}{\mu _i^2})+t_3(1+x_3P^\sigma )\rho ^{1/3}(\frac{\mbold{r}_1+%
\mbold{r}_2}{2})\delta (\mbold{r}), \label{eq31}
\end{equation}
where $P^\sigma $ and $P^\tau $ are the usual spin and isospin exchange
operators, the coefficients $w_i,b_i,h_i$ and $m_i$ are the parameters of
the central force and $\mu_i$ the range of the Gaussian form factor.
The values of these parameters are given in \cite{Gog}.
In DFM calculations the density dependent term is taken in the so-called
''sudden''approximation, which implies that the density of the composite
system is simply the sum of individual densities: $\rho (\mbox{\boldmath$r$}%
)=\rho _1(\mbox{\boldmath$r$})+\rho _2(\mbox{\boldmath$r$}-%
\mbox{\boldmath$R$})$ \cite{JP}.

The density-dependent M3Y force is defined as 
\begin{equation}
v(r)=F(\rho )g(E)v_{M3Y}(r),  \label{eq32}
\end{equation}
where $v_{M3Y}(r)$ is the finite range of the M3Y Paris \cite{Anan}
or Reid-Elliott\cite{Bert}
force with a Yukawa-type formfactor. For a BDM3Y force the density-dependent
term is given by \cite{Kh}:
\begin{equation}
F_{BD}(\rho )=C(1-\alpha \rho ^\gamma ),  \label{eq33}
\end{equation}
while the DDM3Y force \cite{Kob}reads:
\begin{equation}
F_{DD}=C(1-\alpha e^{-\gamma \rho }).  \label{eq34}
\end{equation}
Again, the density-dependent term is considered in the ''sudden'' approximation.
The energy dependent term is taken as $g(E)=1-C_EE$.

In this paper we are going to consider the validity of different
approximations to the DM for the calculation of the DFM potential
rather than obtaining
a realistic potential for describing the elastic scattering data. In order
to do this wee will make the two following simplifications.

The first concerns the way that the ground state density of each isolated $%
^{16}O$ nucleus is calculated. To do this, we minimize the quantal ground
state energy with respect to the oscillator parameter $\varepsilon =\sqrt{m\omega
/\hbar }$. In the semiclassical calculations using the ETF, SL and CB
approaches to the DM we first write the ground state energy using the
Kohn-Sham \cite{KS} scheme (see reference \cite{SV} for more details) and
then minimize with respect to the HO parameter. The binding energy and root mean
square radius (RMSR) of the $^{16}O$ ground state obtained quantally and
semiclassically (with the ETF, SL and CB approaches) are collected in Table 1. One
can see that in all cases the results obtained with the ETF approach to
the DM give a better description of the quantal ground state properties of
$^{16}O$ than that obtained using the semiclassical SL or CB
approximations to the DM. On the other hand, among the considered M3Y
forces, the BDM3Y1$^{*}$(Paris) force gives a RMSR that is very close to
that obtained with the Gogny force and consequently the same ground state
density for the $^{16}O$. Therefore, we will use the Gogny and BDM3Y1$^{*}$%
(Paris) forces in our analysis.

The second remark deals with the definition of relative momentum K. If one
uses the local definition of K and solves the system of equations: V=V(K);
K=K(V) selfconsistently, the nuclear and Coulomb potentials are not
separated. In this paper we are going to consider the different kinds of
forces independently. For this reason and for the sake of simplicity we will
use the global definition of K.

Now we consider the resulting $^{16}O-^{16}O$ ETF potentials calculated
using the quantal HO density for each $^{16}O$ nucleus and the Gogny and
BDM3Y1$^{*}$ forces. In Figure 1 these potentials calculated at several
energies are displayed by dashed and solid lines respectively. The upper
dashed line is the energy independent sum of the Brink-Boeker and zero-range
direct parts of the DFM potential in the Gogny force case. Thus, the
difference between the full potential and this energy independent direct
part gives the contribution of the one-particle exchange due to the finite
(energy dependent) and zero (energy independent) ranges of the Gogny force.
One can see that at low energies the total potential is purely attractive
due to the exchange interaction coming from the finite-range part of the Gogny
force. When the energy increases, the total DF potential tends to its energy
independent part (direct Brink-Boeker and full zero-range contributions).
This latter potential is repulsive at small distances and attractive when
the separation between the nuclei is increased and tends to zero at
large distances.

As can be seen from eq.(\ref{eq32}), the BDM3Y1 force depends linearly
on the energy. This force has two essential differences when
compared with the Gogny interaction. First of all, in this case
the direct DFM potential (%
\ref{eq3}) is energy dependent. The upper solid line
corresponds to this direct potential in the $E=0$ case. On the other hand,
taking into account the linear energy dependence of the BDM3Y force and the
inverse energy dependence of the exchange potential (\ref{eq21}) (in the
high-energy limit) one sees that the exchange potential in this case does not
vanish when the energy tends to infinity and, consequently, the total
potential does not tend to the direct part in this limit. The total and
direct ETF DFM potentials obtained with the BDM3Y1$^{*}$ force are also
displayed in Figure 1. Looking at this figure one can see that in the region
of energies of physical interest ($E_{cm}<1000MeV$) the total DFM potentials
obtained using quantal HO densities and Gogny and BDM3Y1$^{*}$ forces are
very similar. At higher energies they start to show
differences. However, these differences should be considered only in the
mathematical aspect because the nucleus-nucleus potentials at these high
energies can not be defined correctly.

Now we turn to the discussion of the validity of different
approximations to the DM for calculating DFM potentials
considered in the present paper. As was pointed
out above, the exchange effects are larger at small energies and their
contribution falls when the energy increases. Therefore it is reasonable
to expect that the influence of the different approaches to DM on the DFM
potential will be more important at low energies. Hence to investigate the
applicability of the different approximations to the DM
for calculating the DFM
potential, we will consider the static case (K=0). To this aim we define a
function $\sigma $(D), which is the deviation of the approximate total DFM
potential with respect to the exact potential: 
\begin{equation}
\sigma (D)=\frac{V_{SL,CB,ETF}(D)-V_{exact}(D)}{V_{exact}(D)},  \label{eq35}
\end{equation}
where $V_{exact}(D)$ is the DFM potential calculated with the
quantal HO DM at a separation distance $D$, whereas $V_{SL}$, $V_{CB}$ or $%
V_{ETF}$ are the DFM potentials calculated with the SL, CB or ETF
approximations. We perform two sets of calculation using two different
densities for the isolated $^{16}O$ nucleus to obtain the $V_{SL}$,$V_{CB}$
and $V_{ETF}$ for the Gogny and BDM3Y1$^{*}$forces.

In the first case we
used the densities that minimize the ground state Kohn-Sham energy
calculated in the SL, CB or ETF approach for an uncharged $^{16}O$ nucleus.
In other words this case corresponds to the selfconsistent calculation,
where the same approach to the DM is used to obtain the DFM potential
and the density of each isolated nucleus.

These results are plotted in Figures 2 and
3 for the Gogny and BDM3Y1$^{*}$ forces respectively. These figures clearly
show that the DFM potential $V_{ETF}$ calculated using the selfconsistent
ETF densities reproduces better the quantal DFM potential than $V_{SL}$ or $%
V_{CB}$ calculated with the selfconsistent SL or CB densities. This result
is obvious if one takes into account that for an isolated $^{16}O$ nucleus
the selfconsistent density in the ETF approximation is closer to the quantal
density as compared with the other approximation to the DM, as can be seen
from Table 1.

In the second set we obtain the different DFM potentials using the quantal
HO density for each $^{16}O$ nucleus and the Gogny and BDM3Y1$^{*}$
forces. The results for the function $\sigma $(D) are plotted in Figures 4
, 5 and 6 for the Gogny, BDM3Y1$^{*}$ and Coulomb forces respectively.

In this case the situation is more complicated. From Figures 4 and 5 one can
see that $V_{SL}$ (dotted line) does not reproduce the quantal potential
for distances $D>3$fm. The potential $V_{ETF}$ (dashed-dotted line) is very
close to the quantal DFM potential for $D<6$fm. If $D>6$fm, $V_{CB}$
(dashed line) gives the best agreement with the quantal DFM potential,
while for $D<6$fm it works slightly worse than $V_{ETF}$.
At large distances $R\approx $%
10 fm, the nuclear part of the nucleus- nucleus potential is very small wh
compared with its Coulomb part. Consequently, the rather strong deviation (
within 10$\%$) shown by the the nuclear part is not very important.
However, from the formal point of view one should take it into account.

As is discussed in \cite{SV} all the considered approximations to the DM
(SL, ETF or CB) are actually distributions. They are very efficient for
obtaining expectation values, but there is no reason for reproducing the
quantal DM at each point in coordinate space. If one looks for the
asymptotic behaviour of the different approximations to the DM considered in
this paper, one can see that at finite s, the SL and CB DM behave as : $%
\rho (R,s)\propto 2C\alpha ^2R^2\exp [-\alpha ^2R^2]$ when $R\rightarrow
\infty $ (more precisely when $\alpha R>>1,R>>s$), where C is the
normalization constant. The ETF DM has a different asymptotic behaviour: $%
\rho (R,s)\propto 2C\alpha ^2R^2(1-\frac 2{27}\alpha ^4s^2R^2)exp[-\alpha
^2R^2]$. At the same time the quantal DM asymptotically behaves as $\rho
(R,s)\propto 2C\alpha ^2R^2exp[-\alpha ^2R^2]$. One can see that at very
small $s$ and large $R$ the SL, CB and ETF DM are similar to the quantal
density matrix.
However, when $s$ is increased, the ETF DM starts to deviate from the
quantal values. Due to the extra $R^2$ dependence, the ETF DM gives a worse
approximation to the quantal DM as compared with SL and CB DM if one uses the
same quantal HO density for obtaining the different approaches to the
quantal DM.

This lack of accuracy of the ETF DM in describin the quantal DM is not
important when one calculates the ground state properties of isolated
nuclei. In the region of wrong asymptotic behaviou the DM is too small to
give any significant contribution to the total energy as can be seen from
Table 1. However, for calculating DFM potentials the situation is different.
The values of densities over all the nucleus contribute to the DFM potential
at large distances Therefore,the wrong asymptotic part of the ETF DM can
give a significant contribution to the DFM potential at large separaon
distances. Thus, we come to the conclusion that the DFM potential is a more
sensitive tool for investigating the one-particle density matrix in nuclei
than fully integrated quantities such as the binding energy or RMSR.

For improving $V_{ETF}$ calculated with the quantal HO density at large
separation distances $D$, we can make the following modification of the ETF
DM. We introduce a cutoof radius $R_{co}$ that removes the $h^2$ correction
from the SL DM. Therefore this new ETFM DM is is just the ETF DM when $%
R<R_{co}$ and the SL DM if $R>R_{co}$ and has the right asymptotic
behaviour for large R values. The results of the DFM potential calculated
with this new ETFM DM are displayed in Figures 4, 5 and 6 by solid lines. It
is clear that this new approximation gives a better description of the exact
DFM potential at all distances $D$. However, a new parameter, the cutoof
radius in the outer region, is introduced in this case. The solid line
corresponds to the approximated DFM potential obtained with this modified ETF
DM ($V_{ETFM}$), where the $\hbar ^2$ contribution has been cut off after $%
R_{co}$. One can see that $V_{ETFM}$ agrees to within $1\%$ with the
quantal DFM potential.

In Figure 6 we show the results for the long-range Coulomb force. For $D<4$%
fm $V_{SL}$ gives a better result than $V_{CB}$ and at $D>4$fm it starts to
deviate from the t quantal DFM potential values. The $V_{ETF}$ gives a very
good description of the quantal DFM potential and gives an error within 0.25$%
\%$ if $6<D<12$fm. Furthermore, for the Coulomb case
the use of $V_{ETFM}$ significantly improves
the agreement with the quantal DFM potential in all of the
considered region.

It is also interesting to consider in our model calculations the height
and position of the Coulomb barrier. The results corresponding to the
selfconsistent calculations are collected in Table 2. In this case the ETF
DFM potential gives a clearly better approximation to the exact quantal
values when compared with SL and CB DFM approaches.
This result is obvious if one takes into accountthe fact that the
selfconsistent density for the ETF case is very close to the quantal density.
The height and position of the Coulomb barrier calculated with the different
approximations to the DFM potential obtained with the same quantal density
are shown in Table 3. In this case one can see that the better agreement
with the quantal results is obtained using the CB approach to the DFM
potential. To improve the ETF DM potential we again use the ETFM
approxiationto the DM where
a cutoof radius $R_{co}$ is introduced. First of all,from Fig.4-6 one
can obtain the following conclusion:
the longer the range of the force, the smaller the value of cutoof
radius that should be used to reproduce the quantal DFM. The Coulomb barrier
comes from the superposition of longe-range Coulomb force ($R_{co}=5fm)$
and a shorter-range nuclear force. We choose
the cutoof radius $R_{co}=5.5fm$ and
$R_{co}=5fm$ for the Gogny and BDM3Y1$^{*}$ forces respectively.
One can see that the ETFM DFM potential nicely reproduces the
quantal result for both forces. From Table 3 one can also see thatthe
the Gogny and BDM3Y1$^{*}$ interactions give very close results for the
Coulomb barriers.

\section{Summary}

In this paper we have derived the one-particle exchange contribution
to the DFM potential using the recently proposed ETF approach to the DM. We
consider here nucleon-nucleon forces with the widely used Gauss, Yukawa and
Coulomb form factors. As representative examples of these forces, we have
performed our numerical analysis with the Gogny and BDM3Y1$^{*}$ effective
interactions.

To obtain the densities of each isolated $^{16}O$ nucleus, we have used
trial HO wavefunctions. The oscillator parameter $\varepsilon =\sqrt
{m\omega /\hbar}$ is obtained by minimizing the exact ground state energy in
the quantal case and the corresponding Kohn-Sham energy when the quantal DM
is replaced by its SL, CB or ETF approximations.

If each self-consistent density (SL, CB or ETF) is used for obtaining its
corresponding DFM potential, it is found that the ETF potential gives a
better description of the quantal one than the SL and CB DFM potentials.

However, if one uses the same selfconsistent quantal density to obtain all
the approximated DFM potentials, the situation is more complicated. First of
all,it is clearly seen that the SL approach to the DFM potential compares
worse with the quantal potential than the CB or ETF approaches. In all the
analyzed cases the ETF approach to the DFM potential gives a better
description of the quantal potential at small and intermediate separation
distances ($D<6fm$) when compared with the potential obtained in the CB
approximation. However, at large separation distances, the ETF approach
fails in describing the quantal potential. From a physical point of view,
this is not very important because the nuclear part of the total ion-ion
potential is very small at these distances. The reason for this deviation
of the ETF potential from the quantal potential is due to the different asymptotic
behaviour of the quantal and ETF DM of each isolated nucleus.
To improve the ETF DM of each nucleus at large distances,
we drop its $\hbar^2$-contribution after a cutoof radius $%
R_{co}$. In this way one obtains a modified DM which coincides with the
full ETF DM at $R<R_{co}$ and with its SL part DM if $R>R_{co}$. We have
numerically shown that this ETFM DM produces a DFM potential in very good
agreement with the quantal result in the whole range of distances.

Hence we reach the conclusion that the DFM potential is a sensitive tool
for investigating the quality of the approximations to the DM in the whole
range of R-values. While in the calculations of ground state properties of
nuclei the asymptotic behaviour of the DM is suppressed by the small values of
the particle density at large R, in the calculation of the DFM potentials it
is clearly revealed.

\section{Acknowledgments}

One of us (X.V) acknowledges financial support from DGCYT (Spain) under
grant PB95-1249 and from DGR (Catalonia) under grant 1998SGR-00011.

\pagebreak

\section{Table Captions}

Table 1. Binding energy and root mean square radius $<r^2>^{1/2}$ of an
uncharged $^{16}O$ nucleus calculated with trial HO wavefunctions
quantally (QM) and using the SL, CB and ETF approximations to the density
matrix. The results obtained with the Gogny and several M3Y forces are
displayed. The centre-of-mass correction to the energy has been taken
into account. The HO parameter which minimizes the binding energy for each
force is given by $\alpha=3/2 <r^2>^{-1/2}$.

Table 2. Height ($V_B$) and position ($R_B$) of the Coulomb barrier
calculated with the quantal DFM potential (QM) and with its SL, CB and ETF
approximations using the corresponding selfconsistent densities for the
Gogny and BDM3Y1$^{*}$ forces.

Table 3. The same as Table 2 but with the SL, CB and ETF DFM potentials
obtained with the quantal density.

\pagebreak

Figure 1. Direct and exchange contributions at several energies to the
ETF DFM potential calculated with the quantal density.
Solid and dashed lines correspond to the results
obtained with the BDM3Y1$^{*}$ and Gogny forces respectively.

Figure 2. Relative deviation from the quantal DFM potential with respect
to its SL (dotted), CB (dashed) and ETF (solid) approximations obtained
with the corresponding selfconsistent density and the Gogny force.

Figure 3. The same as Figure 2, but for the BDM3Y1$^{*}$ force.

Figure 4. Relative deviation from the quantal DFM potential with respect
to its SL (dotted), CB (dashed), ETF (dashed-dotted) and ETFM (solid)
approximations using the quantal density and the Gogny force.

Figure 5. The same as Figure 4, but for the BDM3Y1$^{*}$ force.

Figure 6. The same as Figure 4, but for the Coulomb force.

\pagebreak

\begin{center}
{\large Table 1}
\end{center}

\begin{tabular}{lllllllll}
& E$_{QM}$ & $<r^2>_{QM}^{1/2}$ & E$_{SL}$ & $<r^2>_{SL}^{1/2}$ & E$_{CB}$ & 
$<r^2>_{CB}^{1/2}$ & E$_{ETF}$ & $<r^2>_{ETF}^{1/2}$ \\ \hline
Gogny & -138.07 & 2.63 & -129.49 & 2.67 & -133.80 & 2.65 & -134.35 & 2.66 \\ 
\hline
BDM3Y1 & -121.15 & 2.55 & -112.67 & 2.57 & -114.63 & 2.57 & -115.76 & 2.56
\\ \hline
BDM3Y2 & -114.57 & 2.53 & -105.54 & 2.55 & -107.82 & 2.54 & -109.21 & 2.53
\\ \hline
BDM3Y3 & -110.69 & 2.52 & -101.46 & 2.53 & -103.86 & 2.53 & -105.33 & 2.52
\\ \hline
BDM3Y1$^{*}$ & -115.57 & 2.63 & -107.94 & 2.65 & -109.70 & 2.65 & -110.44 & 
2.63 \\ \hline
DDM3Y$^{*}$ & -122.29 & 2.67 & -115.17 & 2.70 & -116.69 & 2.70 & -117.13 & 
2.68 \\ \hline
\end{tabular}

\begin{center}
{\large Table 2}

\begin{tabular}{|lllll|}
\hline
& QM & SL & CB & ETF \\ \hline
&  & Gogny &  &  \\ \hline
R$_B(fm)$ & 8.36 & 8.51 & 8.42 & 8.38 \\ \hline
V$_B(MeV)$ & 10.35 & 10.18 & 10.27 & 10.35 \\ \hline
&  & BDM3Y1$^{*}$ &  &  \\ \hline
R$_B$ & 8.46 & 8.57 & 8.50 & 8.43 \\ \hline
V$_B$ & 10.18 & 10.06 & 10.12 & 10.24 \\ \hline
\end{tabular}
\end{center}

\begin{center}
{\large Table 2}

\begin{tabular}{|llllll|}
\hline
& QM & SL & CB & ETF & ETFM(R$_{co}$=5fm) \\ \hline
&  &  & Gogny &  &  \\ \hline
R$_B(fm)$ & 8.36 & 8.40 & 8.35 & 8.35 & 8.37 \\ \hline
V$_B(MeV)$ & 10.35 & 10.31 & 10.36 & 10.38 & 10.35 \\ \hline
&  &  & BDM3Y1$^{*}$ &  &  \\ \hline
R$_B(fm)$ & 8.46 & 8.51 & 8.45 & 8.40 & 8.45 \\ \hline
V$_B(MeV)$ & 10.18 & 10.13 & 10.19 & 10.25 & 10.19 \\ \hline
\end{tabular}
\end{center}
\pagebreak

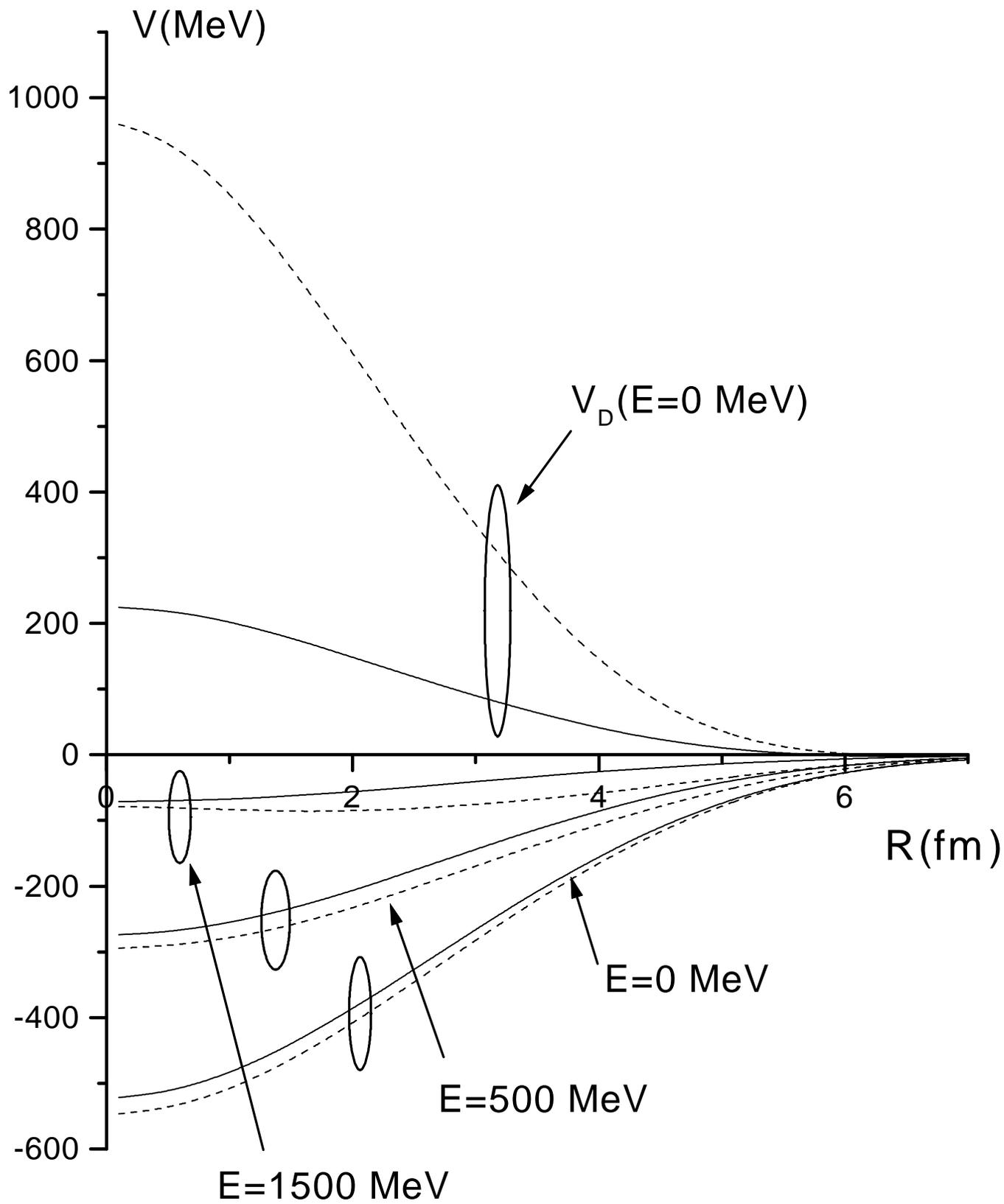
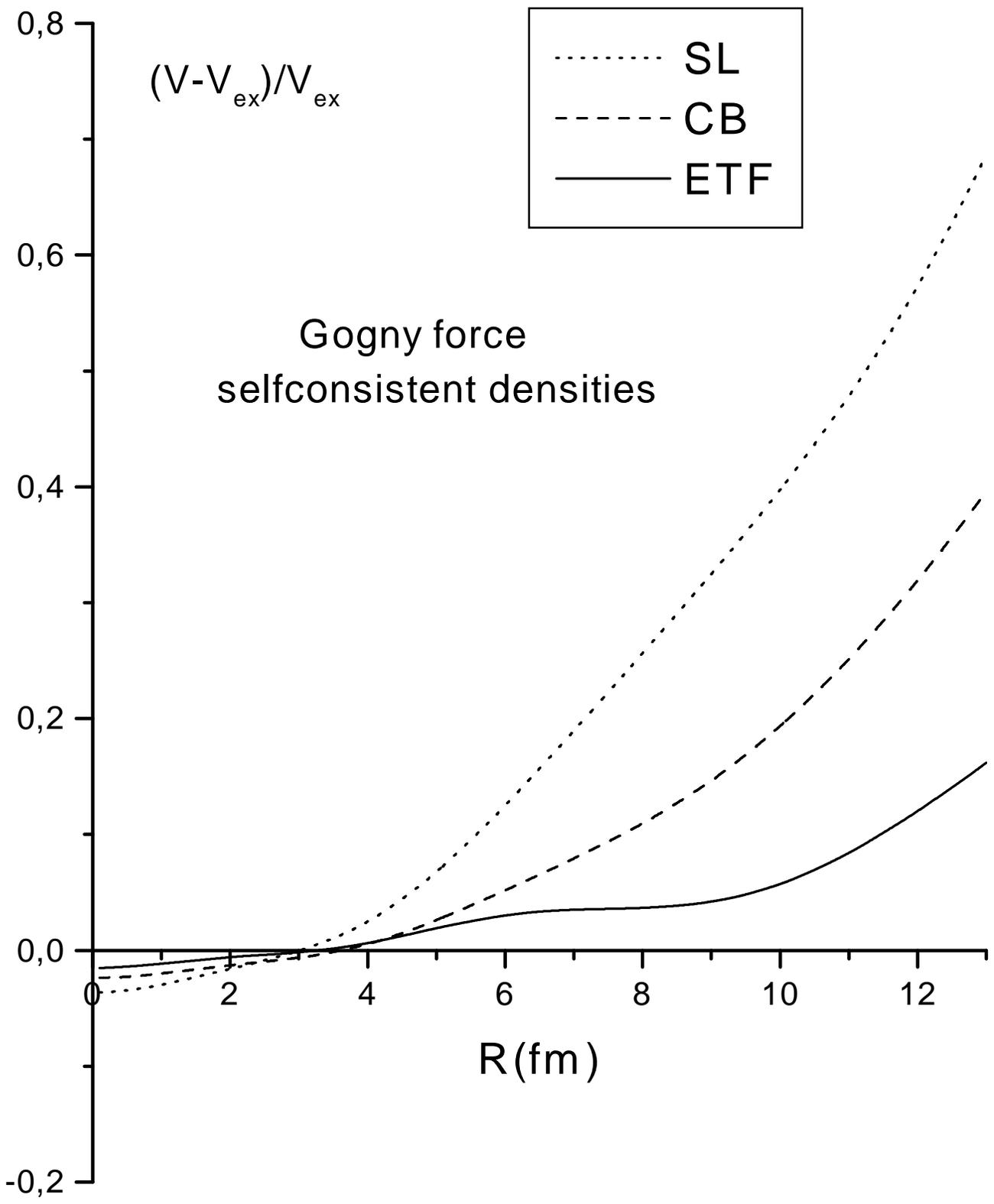
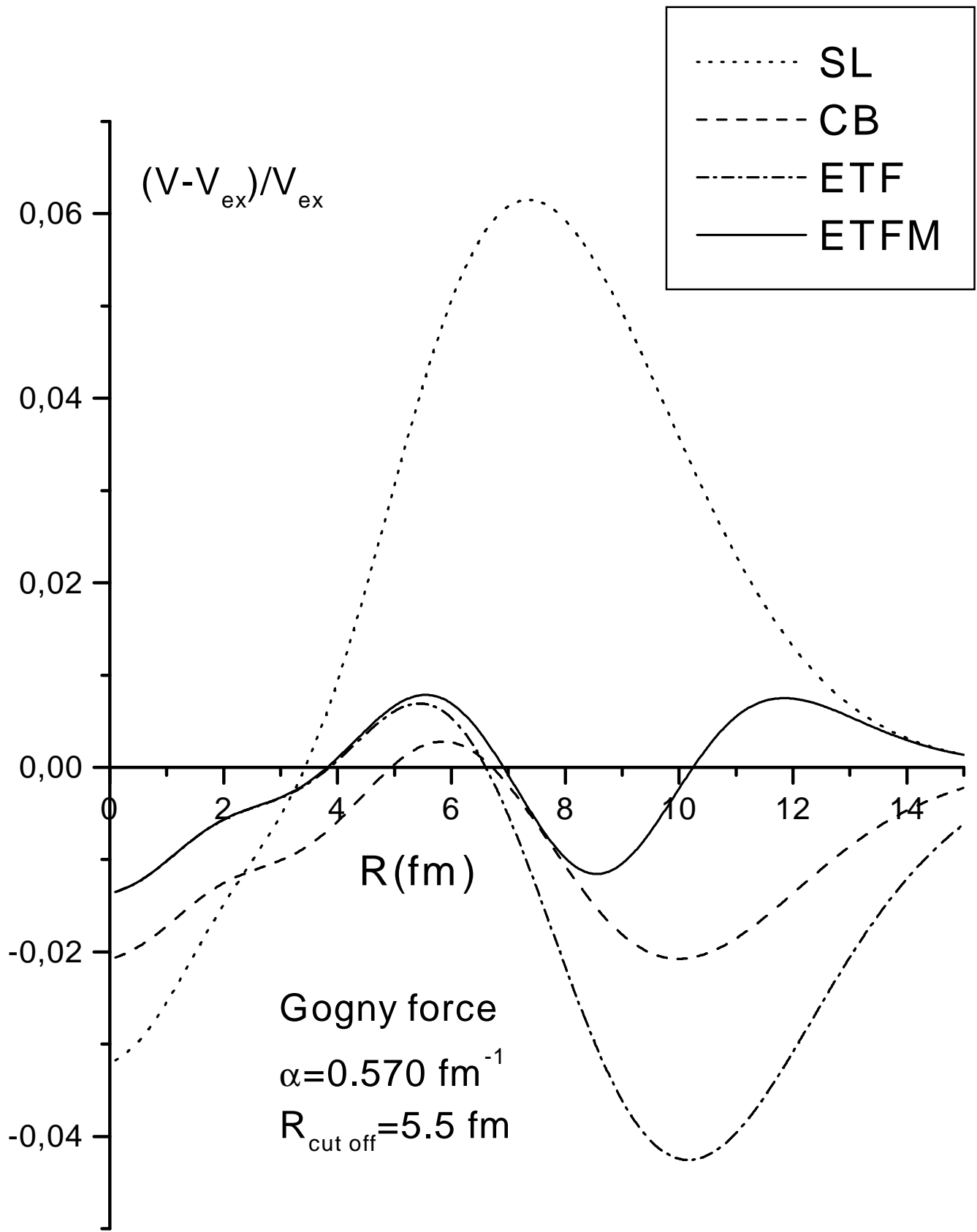
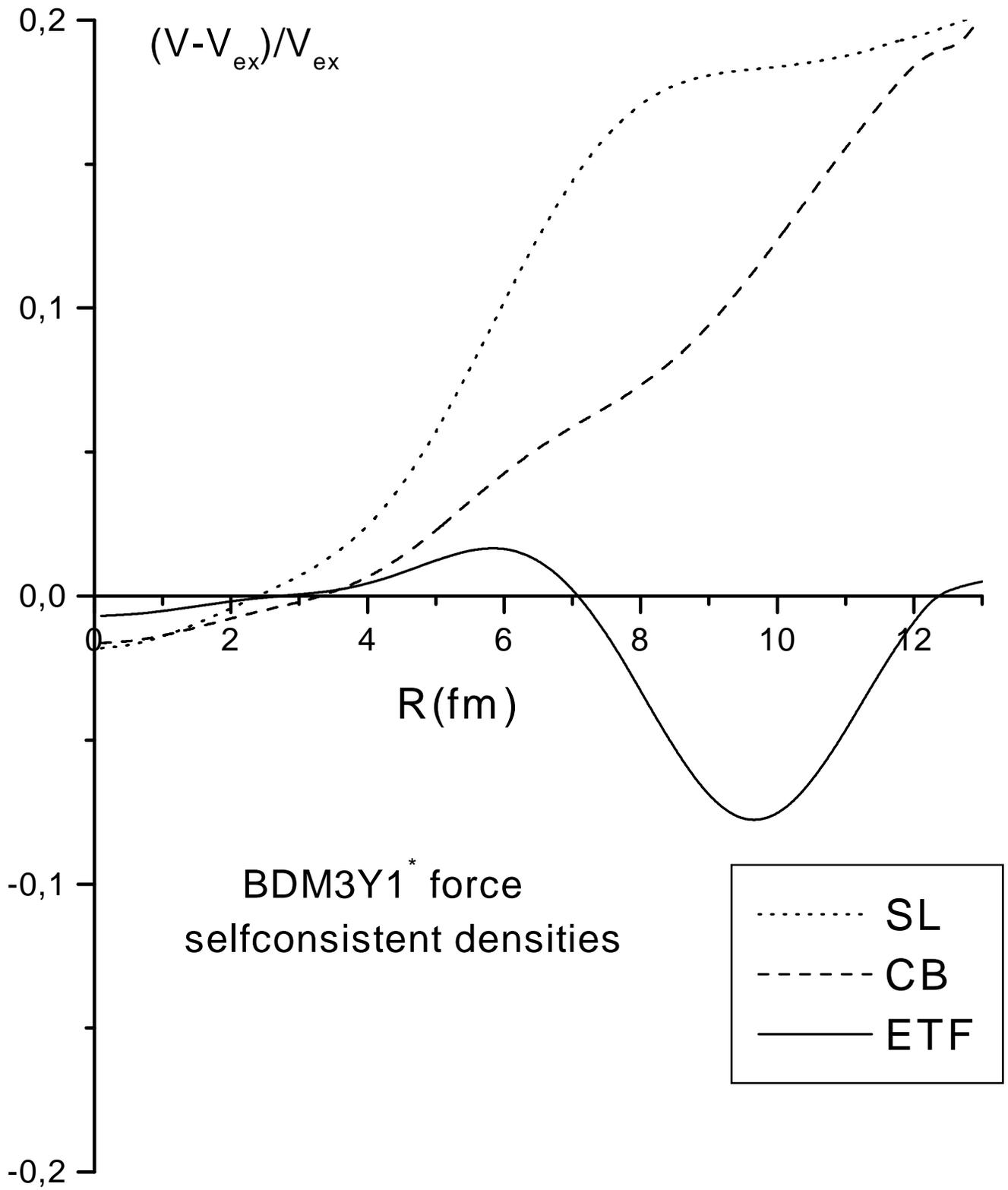
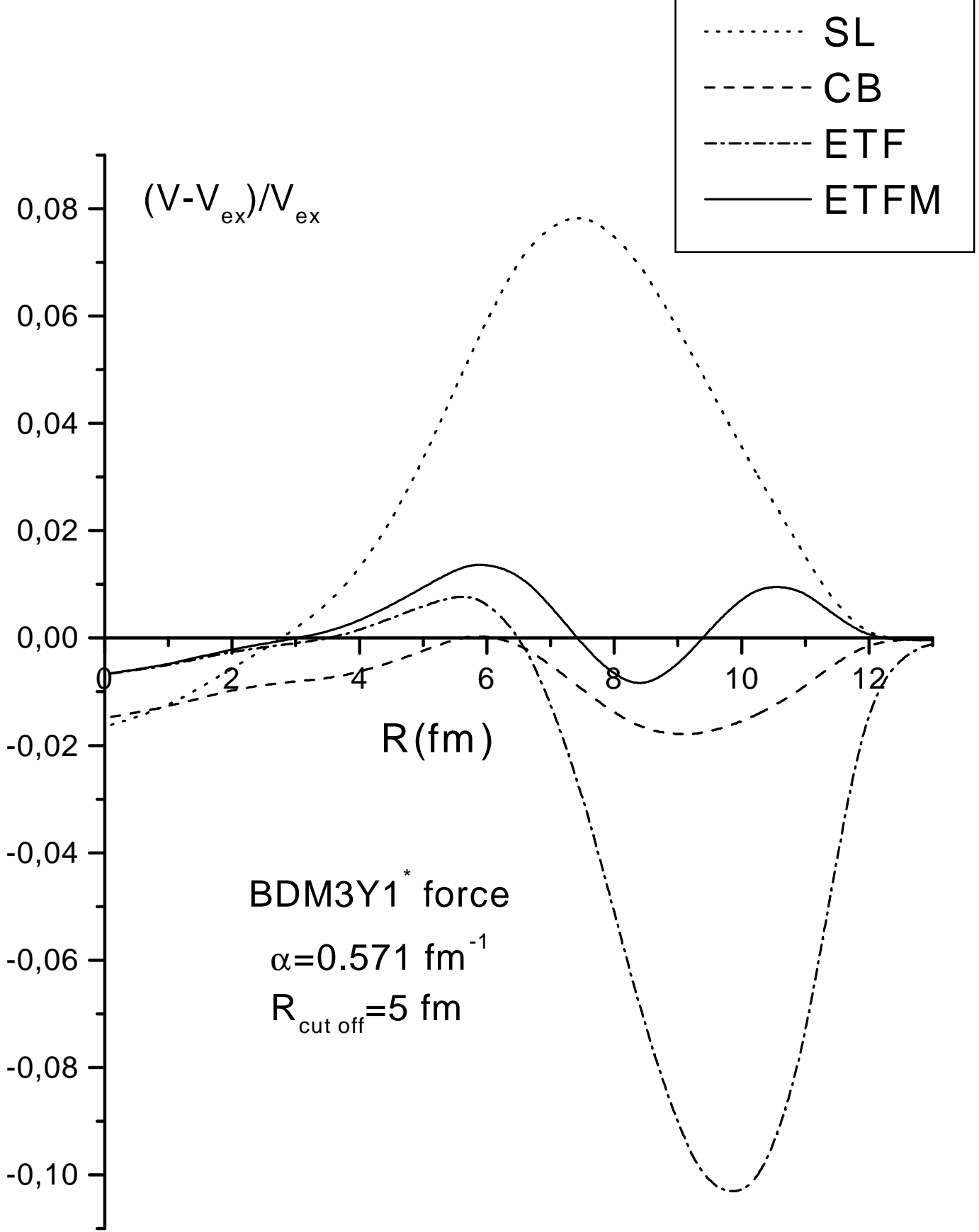
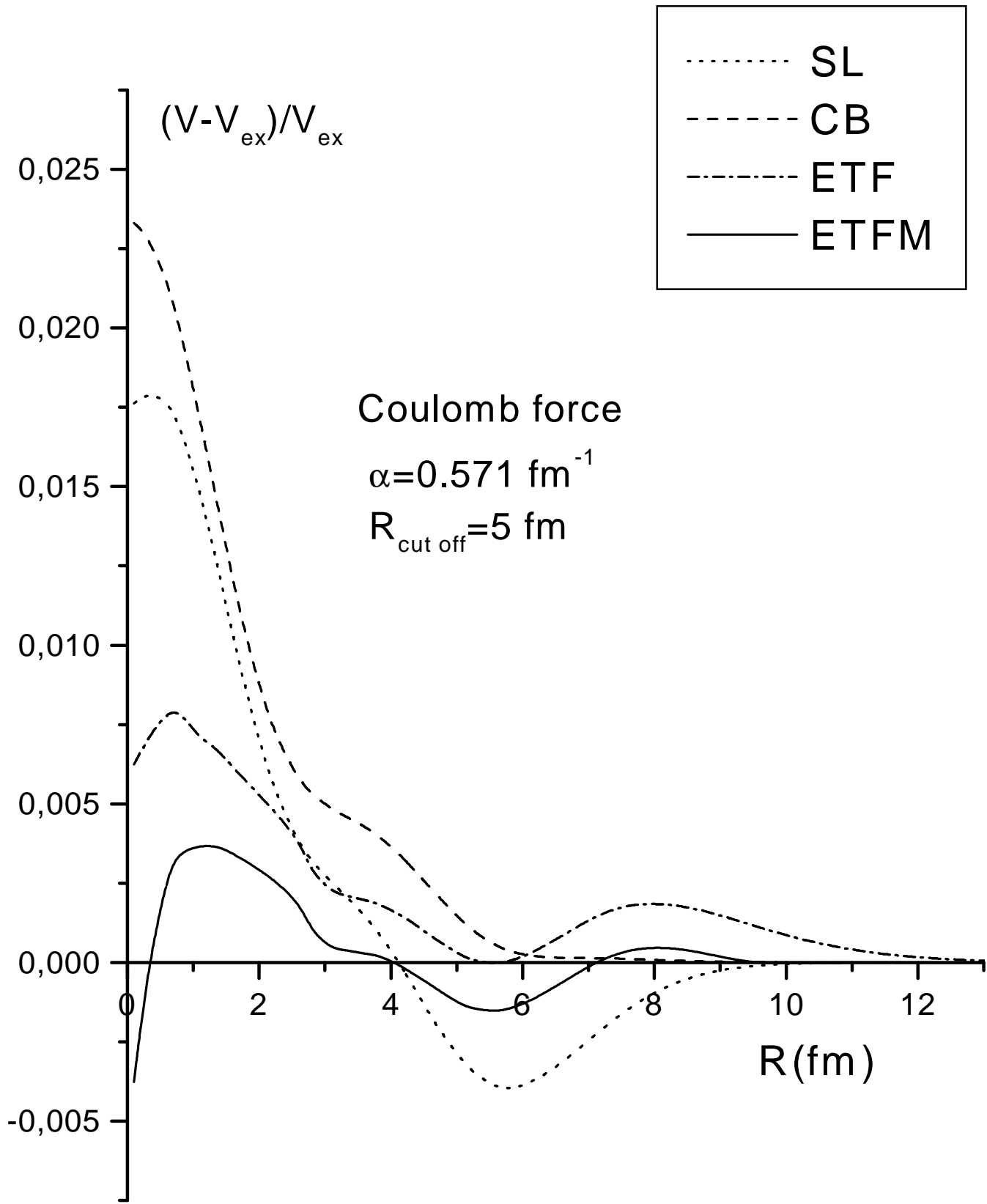
\end{document}